\documentclass[12pt]{iopart}

\begin{document}

\title[Transitions between the Landau levels]{Tomographic probability representation in the problem of transitions between the Landau levels}

\author{E D Zhebrak}

\address{Moscow Institute of Physics and Technology, Institutsky lane 9, Dolgoprudny, Russia}
\ead{el1holstein@phystech.edu}
\begin{abstract}
The problem of moving of a charged particle in electromagnetic field is considered in terms of tomographic probability representation. The coherent and Fock states of a charge moving in varying homogeneous magnetic field are studied in the tomographic probability representation of quantum mechanics. The states are expressed in terms of quantum tomograms. The Fock state tomograms are given in the form of probability distributions described by multivariable Hermite polynomials with time-dependent arguments.\\
The obtained results are generalized in the present work and are applied to determining the transition probabilities between Landau levels.\\
Transition probabilities are calculated using the symplectic tomograms instead of the wave functions. The same method is used in obtaining the transition probabilities between the Landau levels possessed by a charge moving in varying electromagnetic field.

\end{abstract}

\maketitle

\section{Introduction}
In [1] the coherent states of a charge moving in a constant uniform magnetic field were introduced. The coherent states correspond to Gaussian wave packet [2] which moves along the classical cyclotron trajectory with time-independent position dispersions. Such states were also studied in [3] and applied in [4] - [7].
A new formulation of quantum mechanics where the fair tomographic probability distribution was used as alternative of wave function and density matrix has been suggested in [8]. The coherent states and Landau levels of charge moving in the magnetic field were studied in framework of the tomographic probability representation of quantum states in [9]. The charge moving in time-dependent magnetic field was studied in coherent-state representation in [10], [11]. There the parametric excitation of the Landau levels by varying magnetic field was considered by using the coherent state method and explicit expressions for the transition amplitudes were found in terms of classical polynomials. On the other hand the excitation of the Landau levels can be reconsidered in framework of new formulation of quantum mechanics.
The aim of our work is to find the tomographic probability distributions called tomograms describing the charge coherent states and the Landau levels and their nonstationary analogs. We obtain new formulas for transition probabilities between the Landau levels expressed in terms of the tomograms of the  charge quantum states.
The paper is organized as follows. In Sec.2 we review the problem of coherent states both in time-independent and time-dependent magnetic fields. In Sec.3 we construct quantum tomograms of Landau energy level states and the Gaussian packets corresponding to the coherent states. In Sec.4 we provide explicit formulas for the transition probabilities from Landau levels to the ground Landau state expressed in terms of integrals of the state tomograms products.

\section{Coherent states of charge moving in magnetic field}

The problem of a charge moving in magnetic field was studied in [12] for the constant field case and in [9] for the time-dependent case correspondingly.\\
One can consider a charged particle with mass $m=1$ and charge $e=1$ moving in magnetic field $\vec{{\rm H} }=\left(0,0,{\rm H} \right)$ with a vector potential $\vec{A}=\frac{1}{2} \left[\vec{{\rm H} }\times \vec{r}\right]$.  \\ The Hamiltonian of this quantum system will be
\begin{eqnarray}
\fl H =\frac{1}{2} \left[\left(p_{x} -A_{x} \right)^{2} +\left(p_{y} -A_{y} \right)^{2} \right], c=1\label{eq1}
\end{eqnarray}
Let's introduce a cyclotron frequency $\omega \left(t\right)={\rm H} \left(t\right)$. For the constant case it can be introduced as $\omega =1$. Using the method of integrals of motion [1], [10] and introducing the operators
\begin{eqnarray}
\fl\hat{A}^{const} =\frac{\left(p_{x} +ip_{y} \right)+\frac{1}{2} \left(y-ix\right)}{\sqrt{2} }, \nonumber\\
\fl\hat{B}^{const} =\frac{\left(p_{y} +ip_{x} \right)+\frac{1}{2} \left(x-iy\right)}{\sqrt{2} }
\end{eqnarray}
for constant magnetic field and
\begin{eqnarray}
\fl\hat{A}^{var} \left(t\right)=\exp \left(\frac{1}{2} i\int _{0}^{t}\omega \left(\tau \right)d\tau  \right)\frac{\varepsilon \left(t\right)\left(p_{x} +ip_{y} \right)-i\dot{\varepsilon }\left(t\right)\left(y-ix\right)}{2} , \nonumber\\
\fl\hat{B}^{var} \left(t\right)=\exp \left(-\frac{1}{2} i\int _{0}^{t}\omega \left(\tau \right)d\tau  \right)\frac{\varepsilon \left(t\right)\left(p_{y} +ip_{x} \right)-i\dot{\varepsilon }\left(t\right)\left(x-iy\right)}{2} ,\label{eq3}
\end{eqnarray}
 where $\varepsilon \left(t\right)$ as it was stated in [10] corresponds to the both equations $\ddot{\varepsilon }\left(t\right)+\frac{1}{4} \omega \left(t\right)^{2} \varepsilon \left(t\right)=0$ and $\frac{d^{2} }{dt^{2} } \left|\varepsilon \left(t\right)\right|+\frac{1}{4} \omega \left(t\right)^{2} \left|\varepsilon \left(t\right)\right|-\frac{1}{\left|\varepsilon \left(t\right)\right|^{2} } =0$. For axially symmetric time-dependent magnetic field, we can obtain the quantum states corresponding to this motion
\begin{eqnarray}
\label{const}
\fl\left\langle x,y\right. {\left| \alpha ,\beta  \right\rangle} ^{const} =\sqrt{\frac{1}{\pi } } e^{-\frac{\left(x^{2} +y^{2} \right)}{2} } e^{-\frac{1}{2} \left(\left|\alpha \right|^{2} +\left|\beta \right|^{2} \right)+\left[\beta \left(x+iy\right)+i\alpha \left(x-iy\right)\right]-i\alpha \beta }  \label{eq4}
\end{eqnarray}

\begin{eqnarray}
\label{var}
\fl\left\langle x,y\right. {\left| \alpha ,\beta  \right\rangle} ^{var} =\frac{1}{\sqrt{\pi } \varepsilon } \exp \left[\frac{i\dot{\varepsilon }}{2\varepsilon } \left(x^{2} +y^{2} \right)-\frac{1}{2} \left(\left|\alpha \right|^{2} +\left|\beta \right|^{2} \right)+\right.\nonumber\\
\fl\left.\frac{1}{\left|\varepsilon \right|} \left(\beta \varsigma e^{-i\gamma _{-} } +i\alpha \varsigma ^{*} e^{-i\gamma _{+} } -i\alpha \beta e^{-i\left(\gamma _{+} +\gamma _{-} \right)} \right)\right],
\end{eqnarray}
where $\alpha $, $\beta $ are real parameters of coherent state,  $\varsigma =x+iy$ and $\gamma _{\pm } =\int _{0}^{t}\left[\left|\varepsilon \left(\tau \right)\right|^{2} \pm {\rm H} \left(r\right)\right]dr $.\\
In the following consideration of time-dependent  magnetic field we will introduce for brevity $\omega =\omega \left(t\right)$, $ \varepsilon =\varepsilon \left(t\right).$\\
The states (\ref{const}) and (\ref{var}) are called coherent states and they are related to the Fock states as follows
\begin{eqnarray}
\label{fock}
\fl | \alpha ,\beta  \rangle =\exp \left[-\frac{1}{2} \left(\left|\alpha \right|^{2} +\left|\beta \right|^{2} \right)\right]\sum _{n_{1} ,n_{2} =0}^{\infty }\frac{\alpha ^{n_{1} } \beta ^{n_{2} } }{\sqrt{n_{1} !n_{2} !} }  {\left| n_{1} ,n_{2}  \right\rangle}.\
\end{eqnarray}
The corresponding Fock states are eigenstates of the Hamiltonian operator $H$ and angular momentum operator $L_{z} $:
\begin{eqnarray}
H {\left| n_{1} ,n_{2}  \right\rangle} =\left(n_{1} +\frac{1}{2} \right){\left| n_{1} ,n_{2}  \right\rangle} , \nonumber\\
\L_{z} {\left| n_{1} ,n_{2}  \right\rangle} =\left(n_{2} -n_{1} \right){\left| n_{1} ,n_{2}  \right\rangle}.  \label{eq7}
\end{eqnarray}
As it was shown in [1] the motion under consideration corresponds to a Gaussian wave packet with a center moving along the classical trajectory.

\section{Tomographic representation of quantum states and energy levels of a charge moving in magnetic fields }

The function called symplectic tomogram was introduced in [8]. This function connected with density matrix by Radon transform can determine quantum states as well.
\begin{eqnarray}
\label{tgram}
\fl\rm w\left(X,\mu ,\nu \right)=Tr\rho \cdot \delta \left(X-\mu q-\nu p\right),
\end{eqnarray}
where $q$ and $p$ are position and momentum operators respectively, $X$, $\mu $, $\nu $ are reals and $X=\mu q+\nu p$.
The tomogram is nonnegative probability distribution of random variable $X$ which is position in rotated and rescaled reference frame in the phase-space. It is also a homogeneous normalized function. The inverse of (\ref{tgram}) reads
\begin{eqnarray}
\fl\rho =\frac{1}{2\pi } \int \rm w\left(X,\mu ,\nu \right) e^{i\left(X-\mu q-\nu p\right)} dXd\mu d\nu .
\end{eqnarray}
The symplectic tomogram of pure state with the wave function $\psi \left(y\right)$ is determined by the formula:
\begin{eqnarray}
\label{tgram_wf}
\fl\rm w\left(X,\mu ,\nu \right)=\frac{1}{2\pi \left|\nu \right|} \left|\int \psi \left(y\right)e^{\frac{i\mu }{\nu } y^{2} -\frac{iX}{\nu } y} dy \right|^{2} ,
\end{eqnarray}
which is related to fractional Fourier transform of the wave function. Formulas (\ref{tgram})-(\ref{tgram_wf}) can be generalized for a system with several degrees of freedom. For two degrees of freedom the symplectic tomogram $\rm w\left(X_{1} ,\mu _{1} ,\nu _{1} ,X_{2} ,\mu _{2} ,\nu _{2} \right)$is determined by the fractional Fourier transform of the wave function $\psi \left(y_{1} ,y_{2} \right)$ and it reads
\begin{eqnarray}
\label{tgram_two_deg}
\fl\rm w\left(X_{1} ,\mu _{1} ,\nu _{1} ,X_{2} ,\mu _{2} ,\nu _{2} \right)=\frac{1}{4\pi ^{2} \left|\nu _{1} \nu _{2} \right|} \left|\int \psi \left(y_{1} ,y_{2} \right)e^{\frac{i\mu _{1} }{2\nu _{1} } y_{1} ^{2} +\frac{i\mu _{2} }{2\nu _{2} } y_{2} ^{2} -\frac{iX_{1} }{\nu _{1} } y_{1} -\frac{iX_{2} }{\nu _{2} } y_{2} } dy_{1} dy_{2}  \right|.
\end{eqnarray}
For brevity we will use $\rm w=\rm w\left(X_{1} ,\mu _{1} ,\nu _{1} ,X_{2} ,\mu _{2} ,\nu _{2} \right)$. Using $(11)$ we can calculate symplectic tomograms directly from the wave functions. In such a manner there was yielded the symplectic tomogram $\rm w_{\alpha ,\beta }^{const} $ of coherent state of a charged particle moving in a constant magnetic field in [12].
\begin{equation}
\label{tgram_const}
\fl\rm w_{\alpha ,\beta }^{const} =\frac{e^{-\left|\alpha \right|^{2} -\left|\beta \right|^{2} } }{2\pi \sqrt{\frac{\nu _{1} }{4} ^{2} +\mu _{1} ^{2} } \sqrt{\frac{\nu _{2} }{4} ^{2} +\mu _{2} ^{2} } } \left|\exp \left(\frac{\left(\frac{\beta +i\alpha }{\sqrt{2} } -\frac{iX_{1} }{\nu _{1} } \right)^{2} }{1-\frac{2i\mu _{1} }{\nu _{1} } } +\frac{\left(\frac{i\beta +\alpha }{\sqrt{2} } -\frac{iX_{2} }{\nu _{2} } \right)^{2} }{1-\frac{2i\mu _{2} }{\nu _{2} } } -i\alpha \beta \right)\right|^{2}.
\end{equation}
Using the formula (\ref{fock}) we can see that a tomogram of a Fock state $\rm w_{n_{1} n_{2} }^{const} $ can be derived from (\ref{tgram_const}).
\begin{eqnarray}
\fl\rm w_{n_{1} n_{2} }^{const} =\frac{1}{n_{1} !n_{2} !} \frac{1}{2\pi \sqrt{\frac{\nu _{1} }{4} ^{2} +\mu _{1} ^{2} } \sqrt{\frac{\nu _{2} }{4} ^{2} +\mu _{2} ^{2} } } \left|\exp \left(-\frac{X_{1} ^{2} }{\nu _{1} ^{2} \left(1-\frac{2i\mu _{1} }{\nu _{1} } \right)} -\frac{X_{2} ^{2} }{\nu _{2} ^{2} \left(1-\frac{2i\mu _{2} }{\nu _{2} } \right)} \right)\right|^{2}\times\nonumber\\
\fl\times\left|H_{n_{1} n_{2} }^{\left\{S\right\}} \left(\vec{k}\right)\right|^{2}, \
\end{eqnarray}
where $H_{n_{1} n_{2} }^{\left\{S\right\}} \left(\vec{k}\right)$ is an Hermite polynomial of two variables, \\

$S=\left(\begin{array}{cc} {\tilde{b}} & {-i\left(\sqrt{2} \tilde{a}-1\right)} \\ {-i\left(\sqrt{2} \tilde{a}-1\right)} & {-\tilde{b}} \end{array}\right)$, $\vec{k}=\sqrt{2} \left(\begin{array}{c} {\frac{X_{1} }{\nu _{1} -2i\mu _{1} } -\frac{iX_{2} }{\nu _{2} -2i\mu _{2} } } \\ {\frac{-iX_{1} }{\nu _{1} -2i\mu _{1} } +\frac{X_{2} }{\nu _{2} -2i\mu _{2} } } \end{array}\right)$,\\

$\tilde{a}=\frac{1}{1-2\frac{i\mu _{1} }{\nu _{1} } } +\frac{1}{1-2\frac{i\mu _{2} }{\nu _{2} } } $, $\tilde{b}=\frac{1}{1-2\frac{i\mu _{1} }{\nu _{1} } } -\frac{1}{1-2\frac{i\mu _{2} }{\nu _{2} } } $.\\
Analogeously for the time-dependent magnetic field
\begin{eqnarray}
\fl\rm w_{\alpha ,\beta }^{var} =\frac{e^{-\left|\alpha \right|^{2} -\left|\beta \right|^{2} } }{2\pi \left|\left(\frac{\dot{\varepsilon }}{2\varepsilon } \nu _{1} +\mu _{1} \right)\left(\frac{\dot{\varepsilon }}{2\varepsilon } \nu _{2} +\mu _{2} \right)\right|\left|\varepsilon \right|^{2} }\times\nonumber\\
\fl\times\left|\exp \left[-\frac{X_{1} ^{2} }{2\nu _{1} ^{2} \left(i\frac{\dot{\varepsilon }}{\varepsilon } +2i\frac{\mu _{1} }{\nu _{1} } \right)} -\frac{X_{2} ^{2} }{2\nu _{2} ^{2} \left(i\frac{\dot{\varepsilon }}{\varepsilon } +2i\frac{\mu _{2} }{\nu _{2} } \right)} \right]\right|^{2} \left|e^{-\frac{1}{2} \Lambda D\Lambda ^{T} +\Lambda Dl} \right|^{2} ,\
\end{eqnarray}
where $\vec{\Lambda }=\left(\begin{array}{cc} {\alpha } & {\beta } \end{array}\right)$, $D=\left(\begin{array}{cc} {\frac{e^{-2i\gamma _{+} } }{i\left|\varepsilon \right|^{2} } b} & {\frac{e^{-i\left(\gamma _{+} +\gamma _{-} \right)} }{\left|\varepsilon \right|^{2} } \left(i-a\right)} \\ {\frac{e^{-i\left(\gamma _{+} +\gamma _{-} \right)} }{\left|\varepsilon \right|^{2} } \left(i-a\right)} & {-\frac{e^{-2i\gamma _{-} } }{i\left|\varepsilon \right|^{2} } b} \end{array}\right)$, \\

$l=\left(\begin{array}{c} {\frac{e^{-i\gamma _{+} } }{\left|\varepsilon \right|} \left(\frac{X_{1} }{i\left(\frac{\dot{\varepsilon }}{\varepsilon } \nu _{1} +\mu _{1} \right)} -\frac{X_{2} }{\left(\frac{\dot{\varepsilon }}{\varepsilon } \nu _{2} +\mu _{2} \right)} \right)} \\ {\frac{e^{-i\gamma _{-} } }{\left|\varepsilon \right|} \left(-\frac{X_{1} }{\left(\frac{\dot{\varepsilon }}{\varepsilon } \nu _{1} +\mu _{1} \right)} +\frac{X_{2} }{i\left(\frac{\dot{\varepsilon }}{\varepsilon } \nu _{2} +\mu _{2} \right)} \right)} \end{array}\right)$, $a=\frac{1}{\frac{\dot{\varepsilon }}{\varepsilon } +2\frac{\mu _{1} }{\nu _{1} } } +\frac{1}{\frac{\dot{\varepsilon }}{\varepsilon } +2\frac{\mu _{2} }{\nu _{2} } } $, $b=\frac{1}{\frac{\dot{\varepsilon }}{\varepsilon } +2\frac{\mu _{1} }{\nu _{1} } } -\frac{1}{\frac{\dot{\varepsilon }}{\varepsilon } +2\frac{\mu _{2} }{\nu _{2} } } $.
The tomogram of the Fock state can be easily calculated using the property of coherent states to be a generating function for the Fock states:
\begin{eqnarray}
\fl\rm w_{n_{1} n_{2}} ^{var} =\frac{1}{n_{1} !n_{2} !} \frac{1}{2\pi \left|\left(\frac{\dot{\varepsilon }}{2\varepsilon } \nu _{1} +\mu _{1} \right)\left(\frac{\dot{\varepsilon }}{2\varepsilon } \nu _{2} +\mu _{2} \right)\right|\left|\varepsilon \right|^{2} }\times\nonumber\\
\fl\times \left|\exp \left[-\frac{X_{1} ^{2} }{\nu _{1} ^{2} \left(i\frac{\dot{\varepsilon }}{\varepsilon } +2i\frac{\mu _{1} }{\nu _{1} } \right)} -\frac{X_{2} ^{2} }{\nu _{2} ^{2} \left(i\frac{\dot{\varepsilon }}{\varepsilon } +2i\frac{\mu _{2} }{\nu _{2} } \right)} \right]\right|^{2} \left|H_{n_{1} n_{2} } ^{\left\{D\right\}} \left(\vec{l}\right)\right|^{2}.\
\end{eqnarray}

\section{Transition probabilities between Landau levels}

The problem of calculating of transition probabilities between Landau levels with the help of wave function is well studied. Transition probability $\rm P_{n_{1} n_{2} }^{m_{1} m_{2} } $ between the states with the wave functions $\psi _{n_{1} n_{2} } $ and $\psi _{m_{1} m_{2} } $ is equal:
\begin{eqnarray}
\fl\rm P_{n_{1} n_{2} }^{m_{1} m_{2} } =\left|\int \psi ^{*} _{n_{1} n_{2} } \left(x,y\right)\psi _{m_{1} m_{2} } \left(x,y\right)dxdy \right|^{2} ,      \
\end{eqnarray}
where  $\psi _{m_{1} m_{2} } \left(x,y\right)$ correspond to the final and $\psi _{n_{1} n_{2} } \left(x,y\right)$ to the initial state. These two states are non-orthogonal and coherent states can be expanded in them according to the formula $(6)$ for the cases of constant and varying magnetic field as well.
The tomographic approach allows to find $\rm P_{n_{1} n_{2} }^{m_{1} m_{2} } $ in terms of symplectic tomograms:
\begin{eqnarray}
\label{two_states_prob}
\fl\rm P_{n_{1} n_{2} }^{m_{1} m_{2} } =\Tr\rho _{n_{1} n_{2} } \rho _{m_{1} m_{2} } =\nonumber \\
\fl \frac{1}{4\pi ^{2} } \int w_{n_{1} n_{2} } \left(X_{1} ,\mu _{1} ,\nu _{1} ,X_{2} ,\mu _{2} ,\nu _{2} \right) w_{m_{1} m_{2} } \left(Y_{1} ,\mu _{1} ,\nu _{1} ,Y_{2} ,\mu _{2} ,\nu _{2} \right)\times\nonumber \\
\fl\times\e^{i\left(X_{1} -Y_{1} +X_{2} -Y_{2} \right)} dX_{1} dY_{1} d\mu _{1} d\nu _{1} dX_{2} dY_{2} d\mu _{2} d\nu _{2} .
\end{eqnarray}
We consider a situation when the particle possessing a quantum state ${\left| n_{1} ,n_{2}  \right\rangle} $ in a constant magnetic field transit to the ground state ${\left| 0,0 \right\rangle} $ when the time-dependent magnetic field is ``switched off''. In [10] were calculated transition probabilities in terms of wave functions:
\begin{eqnarray}
\fl\rm P_{n_{1} n_{2} }^{m_{1} m_{2} } =\frac{m_{2} !n_{1} !}{m_{1} !n_{2} !} R^{m_{1} -n_{1} } \left(1-R\right)^{n_{2} -n_{1} +1} \left|J_{n_{1} } ^{\left(m_{1} -n_{1} ,n_{2} -n_{1} \right)} \left(1-2R\right)\right|^{2} ,\
\end{eqnarray}
where  $J_{n}^{\left(s,m\right)} \left(x\right)$ is Jacobi polynomial and $R$ can be treated as a reflection coefficient of a particle from the one-dimensional effective potential [10].
Using the symplectic tomograms derived in the Sec. 3 we provide corresponding transition probabilities by the formula (\ref{two_states_prob}):
\begin{eqnarray}
\label{ground_state_prob}
\fl\rm P_{n_{1} n_{2} }^{00} =\frac{1}{16\pi ^{4} } \frac{1}{n_{1} !n_{2} !} \int \frac{1}{\sqrt{\left(\frac{\nu _{1} ^{2} }{4} +\mu _{1} ^{2} \right)\left(\frac{\nu _{2} ^{2} }{4} +\mu _{2} ^{2} \right)} \left|\left(\frac{\dot{\varepsilon }}{2\varepsilon } \nu _{1} +\mu _{1} \right)\left(\frac{\dot{\varepsilon }}{2\varepsilon } \nu _{2} +\mu _{2} \right)\right|\left|\varepsilon \right|^{2} }  \times \nonumber \\
\fl\times \left|\exp \left[-\frac{X_{1} ^{2} }{\nu _{1} ^{2} \left(i\frac{\dot{\varepsilon }}{\varepsilon } +2i\frac{\mu _{1} }{\nu _{1} } \right)} -\frac{X_{2} ^{2} }{\nu _{2} ^{2} \left(i\frac{\dot{\varepsilon }}{\varepsilon } +2i\frac{\mu _{2} }{\nu _{2} } \right)} -\frac{Y_{1} ^{2} }{\nu _{1} ^{2} \left(1+2i\frac{\mu _{1} }{\nu _{1} } \right)} -\frac{Y_{2} ^{2} }{\nu _{2} ^{2} \left(1+2i\frac{\mu _{2} }{\nu _{2} } \right)} \right]\right|^{2} \times  \nonumber\\
\fl\times \left|H^{\left\{S\right\}} _{n_{1} n_{2} } \left(\vec{k}\right)\right|^{2} e^{i\left(X_{1} -Y_{1} +X_{2} -Y_{2} \right)} dX_{1} dY_{1} d\mu _{1} d\nu _{1} dX_{2} dY_{2} d\mu _{2} d\nu _{2} ,
\end{eqnarray}
\\
The integral (\ref{ground_state_prob}) has such a form that the time-dependent function $\varepsilon \left(t\right)$ disappears after integration. For the case of transition between the ground states we have $\rm P_{0,0}^{0,0} =1-R$ [10]. Comparing with (\ref{ground_state_prob}) we obtain an integral equation for reflection coefficient $R$, that can't be calculated directly:
\begin{eqnarray}
\fl\rm R =1-\frac{1}{16\pi ^{4} } \int \frac{1}{\sqrt{\left(\frac{\nu _{1} ^{2} }{4} +\mu _{1} ^{2} \right)\left(\frac{\nu _{2} ^{2} }{4} +\mu _{2} ^{2} \right)} \left|\left(\frac{\dot{\varepsilon }}{2\varepsilon } \nu _{1} +\mu _{1} \right)\left(\frac{\dot{\varepsilon }}{2\varepsilon } \nu _{2} +\mu _{2} \right)\right|\left|\varepsilon \right|^{2} }  \times \nonumber \\
\fl \times \left|\exp \left[-\frac{X_{1} ^{2} }{\nu _{1} ^{2} \left(i\frac{\dot{\varepsilon }}{\varepsilon } +2i\frac{\mu _{1} }{\nu _{1} } \right)} -\frac{X_{2} ^{2} }{\nu _{2} ^{2} \left(i\frac{\dot{\varepsilon }}{\varepsilon } +2i\frac{\mu _{2} }{\nu _{2} } \right)} -\frac{Y_{1} ^{2} }{\nu _{1} ^{2} \left(1+2i\frac{\mu _{1} }{\nu _{1} } \right)} -\frac{Y_{2} ^{2} }{\nu _{2} ^{2} \left(1+2i\frac{\mu _{2} }{\nu _{2} } \right)} \right]\right|^{2}\times\nonumber \\
\fl\times e^{i\left(X_{1} -Y_{1} +X_{2} -Y_{2} \right)} dX_{1} dY_{1} d\mu _{1} d\nu _{1} dX_{2} dY_{2} d\mu _{2} d\nu _{2} .\
\end{eqnarray}

\section{Conclusion}

To resume we point out the main results of our work. We studied the problem of finding the transition probabilities between the Landau levels induced by time-dependence of the homogeneous magnetic field during some period of time using the tomographic probability description of the quantum states. The transition probability is expressed in terms of nonlocal integral of the product of tomograms of initial quantum state and final quantum state with exponential kernel. We have shown that this integral is expressed in terms of Jacobi polynomial which corresponds to standard calculation of the transition probabilities by using the overlap integral of the initial and final wave functions. We presented explicit expression for the integral given in terms of tomograms for the case of transition probabilities from initial Landau level with arbitrary energy and angular momentum to the ground state Landau level. We generalize the obtained result to the case of presence of the varying electric field in future work.

\section{Acknowledgements}

The author thanks the organizers of the Central European Workshop on Quantum Optics 2012 and Moscow Institute of Physics and Technology for partial support.

\section{References}

\numrefs{12}
\item Malkin and V. I. Man'ko, Zh. Eksp. Teor. Fiz.,V.55, p.1014 (1968) [Sov.Phys.JETP, V.28, p.527 (1969)].
\item E.H. Kennard, Z.Physic, V.44, p.326 (1927).
\item Albert Feldman and A. H. Kahn, Phys. Rev. B, V.1, p.4584 (1970).
\item T. Kuzmenko, K. Kikoin, and Y. Avishai, Phys. Rev. B, V.73, I.23, 235310 (2006).
\item V. V. Belov and V. P. Maslov, Dokl. Akad. Nauk SSSR, V.311, p.849 (1990).
\item E.I. Rashba, L.E. Zhukov and A.L. Efros, Phys.Rev. B, V.55, p. 5306 (1997).
\item V.G. Bagrov, S.P. Gavrilov, D.M. Gitman and K. Gorska, J.Phys.A, V.45, 244008.
\item S.Mancini, V.I.Man'ko, P.Tombesi, Phys.Lett.A, V.213, p.1 (1996).
\item V. I. Man'ko, E. D. Zhebrak, "Tomographic Probability Representation for States of Charge moving in Varying Field" // arXiv:1204.3427v1 [quant-ph].
\item A. Malkin, V. I. Man'ko, and D. A. Trifonov, Phys.Rev.D, V.2, p.1371 (1970).
\item V.Dodonov and V.Man'ko, In: Sov. Phys. Proc. FIAN 183 (1988), p. 71.
\item V.I. Manko, E.D. Zhebrak, Optics and Spectroscopy, V.111, p.666 (2011).

\endnumrefs





\end{document}